# Enhancing Data Security by Making Data Disappear in a P2P Systems

Prashant Pilla

Computer Science Department

Oklahoma State University, Stillwater, OK – 74078

ppilla@okstate.edu

**ABSTRACT**

This paper describes the problem of securing data by making it disappear after some time limit, making it impossible for it to be recovered by an unauthorized party. This method is in response to the need to keep the data secured and to protect the privacy of archived data on the servers, Cloud and Peer-to-Peer architectures. Due to the distributed nature of these architectures, it is impossible to destroy the data completely. So, we store the data by applying encryption and then manage the key, which is easier to do as the key is small and it can be hidden in the DHT (Distributed hash table). Even if the keys in the DHT and the encrypted data were compromised, the data would still be secure. This paper describes existing solutions, points to their limitations and suggests improvements with a new secure architecture. We evaluated and executed this architecture on the Java platform and proved that it is more secure than other architectures.

## 1. INTRODUCTION

The storage in Cloud and in P2P networks is different from that of data stored on personal machines. Data in Cloud and in P2P networks is distributed over many servers and could be compromised at any time if not properly secured.

Trusting Cloud and P2P systems for securing confidential data is risky. A simple solution is to encrypt the data and store it in a database to avoid archiving and caching. But, even after encryption the data can be decrypted by the service provider, because the service provider has access to all the data, keys and it may have been cached. A simple encryption is not enough to keep the data secure.

If the data is available for a fixed time, the possibility of attacking the data can be minimized. But it is difficult to ensure that the data is deleted completely after the timeout, as there might be copies. A solution to this problem is to store data in an encrypted form and delete the key after the timeout. It is easier to manage keys rather than the data.

Our main objective is to protect data from malicious, legal, and illegal attackers. R. Geambasu, T. Kohno, A. A. Levy and H. M. Levy from the University of Washington have proposed the Vanish system [1] that follows the timeout concept for keeping the data secure in the Cloud. The objective of the Vanish system is to create data that self-destructs or simply vanishes after a fixed time, and it occurs without any external actions, so that the data is no longer accessible to anyone, even to the user. Initially the Vanish system encrypts the data using a random key $K$ and then uses Shamir's secret sharing technique [2] to split the key into $n$ shares, where collecting $k$ (threshold) shares can reconstruct the original key. Vanish later stores these random $n$ shares in the DHT (Distributed hash table) [3].

The DHT takes on a central role in making data disappear, where the data in the nodes of the DHT will naturally disappear due to the churn effect, arising out of new nodes joining and old nodes leaving the network dynamically over time, making it impossible to determine which node is responsible for storing data. It is possible to construct the original key or VDO (Vanish Data Object) if the Vanish system collects at least *k* of *n* shares before the time expires. If the time expires, the nodes will automatically vanish making it impossible to extract the *k* shares, and hence no one can reconstruct the original key, making the data inaccessible.

Although it provides many advantages, Vanish has shortcomings, which are described below:
1. The minimum timeout is fixed
2. In order to extend the lifetime of a VDO, Vanish system extracts the key shares, generates another set of key shares and distributes them to the nodes.
3. Key shares are available at the node after the timeout.
4. Sybil attacks [4]: Sybil attacks works by continuously crawling the DHT and saving each storage value before timeout.
5. The key shares can be recognized by their key length.

Based on the above shortcomings, we construct a secure Cloud or P2P storage service on top of a public DHT and server infrastructure, where we store the encrypted data in the database and then delete the keys after a specific timeout mentioned by the user. In this architecture we apply centralized and decentralized techniques to make the system more robust against powerful adversaries. We apply the decentralizing technique by using the DHT, and the centralized technique using the Ephemerizer [5], which is a central server. Key K is a combination of two keys: one is stored on the central server Ephemerizer, and the other is stored in the DHT. Similarly, in order to reconstruct original key K for decryption, the system has to gather both of the keys, one from the Ephemerizer server and the other from the DHT. This centralized and decentralized approach makes the system more robust, as they share risks and threats. This self-destructive data is best used in email systems and in Cloud storage, and IAAS (infrastructure as a service). Using this system, the data can be protected against legal attackers.

In order to implement this architecture, we use encryption techniques like AES [6], recursive secret sharing [7], [8], [9], [10], and hashing [11] and perform key management techniques using the Open Chord DHT [12], and global scale P2P.

This architecture is capable of keeping data secure and destroys it after time out even:
- If attacker has a copy of the data and keys.
- Without any explicit delete action by user.
- Without modifying the data or keys.

Our method uses Vanish system as the base architecture but we modify it to improve performance and security to avoid low cost Sybil attacks.

### 1.1 Vanish System

It is known that the threat for a data can be reduced if data was available for a limited time. But it is impossible to remove data completely. A simple solution for this problem is to encrypt data before saving it on the database or server and then control the lifetime of the decryption key. Keys are small in size they are easy to manage than the huge data.

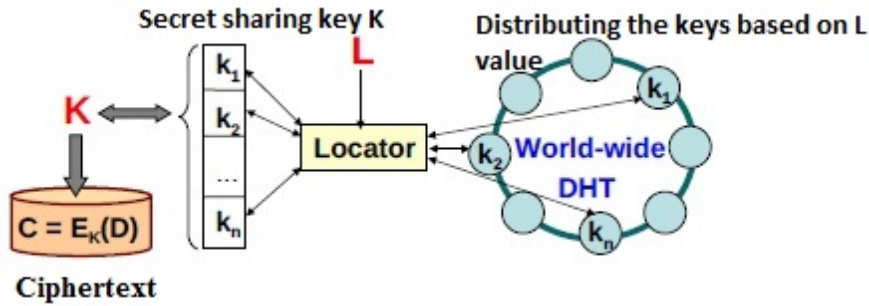

Figure 1: Vanish system

In the Vanish system, keys are placed in the nodes of DHT as shown in figure 1. These nodes have a tendency of disappearing after a period of time, leading to disappearance of decryption keys. Vanish system can be described in two steps, encapsulation, and decapsulation.

**Encapsulation:**

Alice wishes to send a data to Bob.
1. Alice makes a data object $D$ and encapsulates it into a VDO (Vanish data object) $V$. In order to encapsulate $D$ Alice selects a random key $K$ and encrypts the $D$ to obtain a cipher text $C$. (Figure 2)
2. Later, Alice uses the threshold secret sharing and splits the data key $K$ into $n$ pieces of shares $K_1, K_2, K_3...K_n$.
3. Once Alice has computed the key shares $K_1, K_2, K_3....K_n$ she picks a random access key $I$, and selects a pseudorandom number generator keyed by $L$ and derives $n$ indices into the DHT $I_1,I_2,I_3...I_n$.
4. Alice then sprinkles the $n$ shares $K_1, K_2, K_3...K_n$ at these pseudorandom location throughout the DHT, specifically for each $I$ belongs to $\{1,2,..n\}$.
5. Alice stores the shares $K_i$ at the index $I_i$ in the DHT.
6. Finally Alice sends a VDO $V$ consists of *(L,C,n,threshold)* where
   $L$: Key to derive the indices related to that $C$.
   $C$: Cipher text.
   $n$: number of partitions or shares.
   *threshold*: minimum number of shares to create the key $K$.

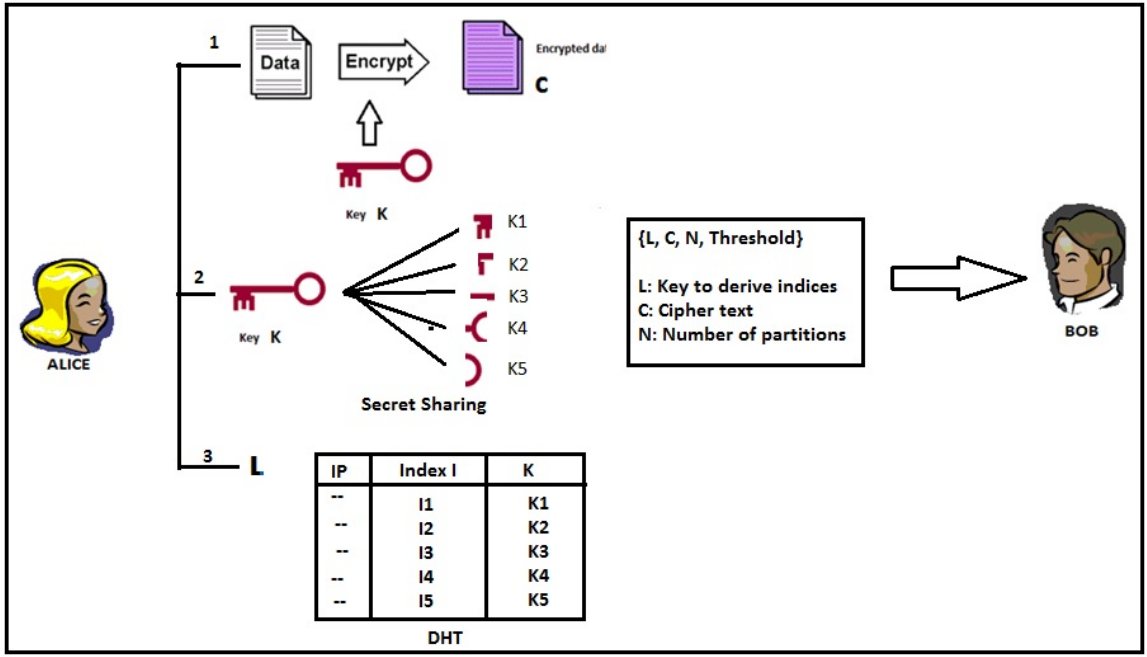

Figure 2: Vanish encapsulation

**Decapsulation:**

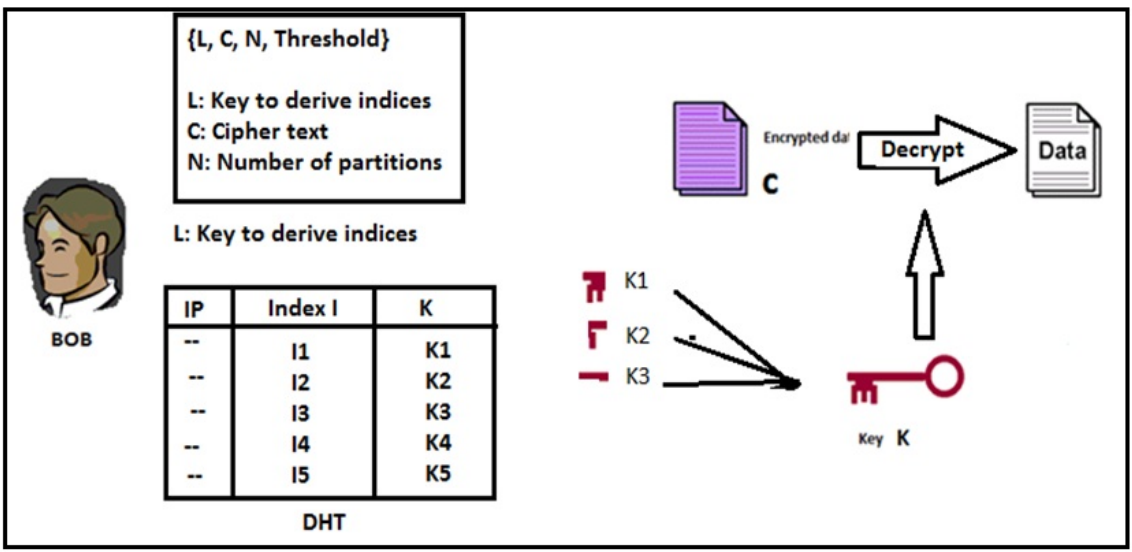

Figure 3: Vanish decapsulation

1. On receiving VDO, Bob extracts the access key $L$ and derives the location of the shares. (Figure 3)
2. Bob retrieves the threshold number of shares to reconstruct the key $K$ and decrypts the cipher text $C$ to get the data $D$.

The main objective of Vanish is to secure the data by not deleting the original data, but by making keys disappear in DHT. Vanish was released in August 2009 and it consists of encapsulation and decapsulation functions, and a firefox plugin [14]. The user selects the text data and converts it into a VDO by right clicking it and selecting the VDO option. It breaks encryption keys into 10 shares with a threshold of 7. These shares are pushed into Vuze DHT [13], which has a default time out of eight hours. After 8 hours, keys are vanished unless the keys are periodically reposted in DHT.

## 1.2 Ephemerizer:

This system also uses same principle as that of Vanish, keeping data available for a finite time and making it unrecoverable after a specific time. In this system, instead of distributing the decryption keys into the DHT, an external server is used to store the decryption key with their time out. During decryption the server provides user with decryption keys till timeout. The Ephemerizer server takes care of key management tasks like key creation, advertising, and deletion. It creates keys for encryption and stores them on the server and checks them periodically for their timeout. It also sends keys for decryption only if their time is not expired.

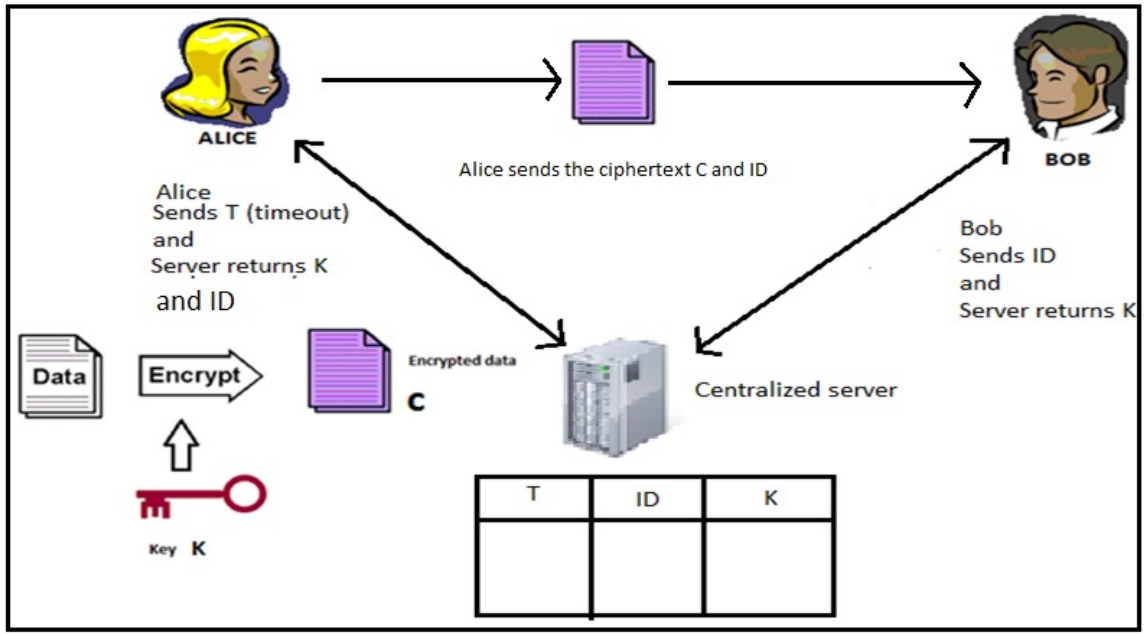

Figure 4: Ephemerizer architecture

This system can be explained in detail using an example as shown below:

Alice wishes to send a data to Bob.

1. Alice sends a request to the Ephemerizer *server* for a key $K$ which will be used to encrypt the data $D$ given the expiration time $T$. (Figure 4)
2. The Ephemerizer generated a random secrets key $K$, ID for that request.
3. The Ephemerizer sends *(K, ID)* to Alice and stores the key $K$ and the timeout $T$ in its database.

4. Alice encrypts the data *D* with the key *K*, and sends (*C, ID*) to Bob.
5. On receiving Bob requests the Ephemerizer for the decryption key *K* for that particular *ID*.
6. The Ephemerizer searches its database for the secret key *K* for that particular *ID* and checks its timeout *T*, if expired return an error message to Bob else return *K*.
7. If valid Bob decrypts the cypher text *C* and extracts the data *D*.

## 1.3 Recursive Secret Sharing:

A recursive secret sharing is like an ordinary secret sharing system except that it makes it possible to store additional secrets. The recursive computational multi secret sharing technique hides *k - 2* secrets of size *b* each into *n* shares of a single secret *S* of size *b*, such that any *k* of the *n* shares sufficient to recreate the secret *S*. It also makes it possible to store additional secret messages $s_1, s_2, s_3$... along with the main secret *S*, so that on combining the shares $K_1, K_2, K_3...K_k$ of the key *K* the main secret *S* and the additional secret messages $s_1, s_2, s_3$... can be extracted. This technique does not require any encryption key. The recursive secret sharing is mainly used for:

- Hiding additional secrets, which are recursively stored within the shares
- Validation and authentication of shares during the reconstruction phase, which provides cheating detection. The hidden secret share is usually checked for authentication.

The recursive secret sharing algorithm is as follows. There are two phases: the dealing phase – the process of making shares with the secret and hidden information; and the reconstruction phase – the process of combing the shares producing secrets and hidden information.

**Dealing Phase**
1. Consider *k - 2* secrets $s_i \in Z_p$, $1 \leq i \leq (k - 2)$.
2. Choose prime $p = max(s_i, S)$, for all $1 \leq i \leq (k - 2)$.
3. Randomly and uniformly choose a number $y_{11} \in Z_p$ and map it as point *(1, $y_{11}$)*.
4. Do for $1 \leq i \leq (k - 2)$
   a) Interpolate points *(0, $s_i$)* and *(j, $y_{ij}$)*, for all $1 \leq j \leq i$ to generate a $i^{th}$ degree polynomial $p_i(x)$.
   b) Sample the polynomial $p_i(x)$ at *i + 1* points: $y_{(i+1)j} = p_i(j + i)$, for all $1 \leq j \leq (i + 1)$.
   c) Map the *i + 1* points as: *(j; $y_{(i+1)j}$)*, for all $1 \leq j \leq (i + 1)$.
5. Interpolate points *(0, S)* and *(j, $y_{(k-1)j}$)*, for all $1 \leq i \leq (k - 1)$ to generate $(k-1)^{th}$ degree polynomial $p_{k-1}(x)$.
6. Sample $p_{k-1}(x)$ at *n* points to generate *n* shares :*(i, $p_{k-1}(i)$)*, for all $k \leq i \leq k + n - 1$.

**Reconstruction Phase**
1) Interpolate any *k* shares to generate $(k - 1)^{th}$ degree polynomial
   $p_{k-1}(x) = S + a_1 x + a_2(x^2) + ..... + a_{k-1} x^{k-1}$.
2) Evaluate $S = p_{k-1}(0)$.
3) Do for $i = k - 2$ down to 1
   a) Map the coefficients of polynomial $p_i(x)$ as points:
   *(j, $a_j$)*, for all $(i + 1) \leq j \leq 2(i + 1)$.
   b) Interpolate *(j, $a_j$)*, for all $(i + 1) \leq j \leq 2(i + 1)$, to generate polynomial $p_i(x)$ of degree *i*.
   c) Evaluate $s_i = p_i(0)$.

## 1.4 DHT:

A DHT distributes data over a large P2P network, so that we can quickly find any given item and distribute responsibility for data storage. The design of a DHT varies like Apache Cassandra, BitTorrent DHT, CAN, Chord, Kademlia, Pastry. The basic operations of a DHT are *Store(key; val)*, *val = Retrieve(key)*, where a key controls which node(s) stores the value *val*, and each node is responsible for some section of the space. Vanish stores keys in Vuze DHT nodes which consists of million nodes, and it is modified Kademlia [15] DHT. Every key and node is assigned to an ID of 160 bit, where each key is stored in the node whose ID is closer to its key ID. Although *Store*, *Retrieve* are the basic operations, the root principle operation that guides these functions is the *lookup* operation. *Lookup* operation searches for the node that holds a specific ID. In Kademlia each node will lookup 20 other closest nodes for the ID it is searching for.

To store or retrieve a key, the requesting node hashes the key to get its ID of the key. This ID is used to map key to the desired node. The requesting node sends a request to 20 other nodes that are close to the ID. After finding node closest to the ID, the requesting node contacts it directly. If a node wants to join the Vuze DHT, it contacts a peer it knows and requests for a lookup for its own ID. After the lookup it finds nodes that are close to its ID. When this new node contacts the new peer with the ID among the 20 closest, they replicate all the stored keys to that node.

## 2. PROBLEM DESCRIPTION

The main drawback of the Ephemerizer architecture is that it is centralized. This third party may not be trustworthy, as the Ephemerizer may still keep some copies of the keys in its cache memory. This calls for a decentralized approach with fewer risks.
The decentralized Vanish approach has drawbacks

1. The minimum timeout is fixed.
   For Vuze architecture, the timeout is 8 hours and for OpenDHT [16] it is about a week. Timeout varies with different DHT architectures. So the data or the key shares remain active in the node for a minimum of 8 hours. This timeout window is sufficient for attackers to steal the keys before they expire. what if a user wants it for one hour?
2. If a user wants the timeout to be of 20 hours, Vanish extend the life of a VDO by a refresh mechanism. The refresh mechanism retrieves original data key *K* from the DHT before its timeout and re-splits it, obtain a fresh new shares and derives a new DHT indices *I1, I2,..IN* and redistributes them in the DHT. The cost of this operation is high, as the Vanish has to decapsulate the shares and
   generate a new shares and encapsulate them again. In order to perform this action, periodic internet connectivity to a PC is required which is not possible for the users who are mobile.
3. Even after 8 hours (for Vuze) or the timeout the key shares are still available at the node. Only the IP of the node changes, but the data in them remains the same. These nodes may reappear with a different IP but they hold the same share.

4. The Key shares can be recognized by their key length. As per the Vanish design code the default key is a 128-bit encryption key and all shares holds the same size, which can be easily recognized and cached.
5. Sybil attacks: Sybil attacks work by continuously crawling the DHT and saving each storage value before timeout.

Vanish utilizes public DHT Vuze Bit-Torrent. The Vuze DHT clients periodically replicates keys they store to other peers that are close to it in order to extend the life of a VDO. Each Vuze client manages a routing table that categorizes peers into a number of "K-buckets" by their distance from its own ID.
The replication properties of the Vuze make the Sybil attack much easier.
1. To increase availability, Vuze replicates the keys to new clients as soon as they join the network.
2. To ensure resiliency as node joins and leave, Vuze node replicates data they know to their neighbors at frequent intervals, usually for every 30 minutes.

A node joins the Vuze DHT by contacting a known peer and starting a lookup for its own ID. It uses a lookup to build its own list of peers and eventually find node that is closest to its ID.

Using the above replication properties, Sybil node takes very less time to gain majority of keys active at that time. It filters necessary to unnecessary keys by measuring its size. A Sybil simply hops from one IP to another, through the available identities, thus gaining almost all keys stored at different locations. Vuze considers these new fake nodes which act as a new natural node and provides them with replicated keys. It takes 3 minutes to hop from one IP to another. For an 8 hour period the each Sybil can hop 160 node ID's with a minimal loss in coverage. Sybil attack is optimized by considering default key share size. The Vanish architecture has a default share of 128 bit encryption key. Sybil considers only those keys with the default size fixed by the Vanish code and leaves rest of the keys or data that does not match default key length.

Even in decentralized approach privacy of data is not guaranteed. It is unsafe for data or keys to be handled by only one source (Ephemerizer) or by everyone (DHT). In centralized Ephemerizer approach the control or the secret is handled by a third party holder, who is considered to be untrustworthy. In the decentralized Vanish approach, it distributes the secret to public DHT which is also unreliable as data is placed in public nodes. This decentralized approach gives many opening for hackers to steal, cache or record the keys and use them to decrypt the encrypted data.

### 3. PROPOSED ARCHITECTURE

In order to overcome the drawbacks, we require a new secure architecture where data is secure and control over the keys are neither completely under centralized nor decentralized sources. The responsibility over the keys is managed by centralized as well as decentralized sources. Sharing responsibilities over the keys will help in building a more secure architecture where data can be secure till expiration.

In these new architecture we use centralized and decentralized approach along with a new secret sharing method, which is recusrsive secret sharing. A third party server is introduced in this architecture, which is used to create, advertise and destroy keys and a public DHT is used which also distributes keys to public nodes. Types of encryption that occurs in this method are data encryption and secret sharing encryption. Data encryption is applied to data and secret sharing encryption is applied to secret $N_t$. The new architecture can be explained in two steps, encapsulation and decapsulation.

### 3.1 Encapsulation:

Alice wants to send a data or message to Bob.

1. Alice sends a request to Ephemerizer server for a key $H$ to encrypt data into cyphertext $C$ for a timeout T. (Figure 5)
2. Ephemerizer stores timeout T and generates a random nonce $N_t$, an ID $ID_t$ and a secret $S_t$ for that request. Then calculates hash $H = (N_t, S_t)$ and stores $\{T, S_t, ID_t\}$.

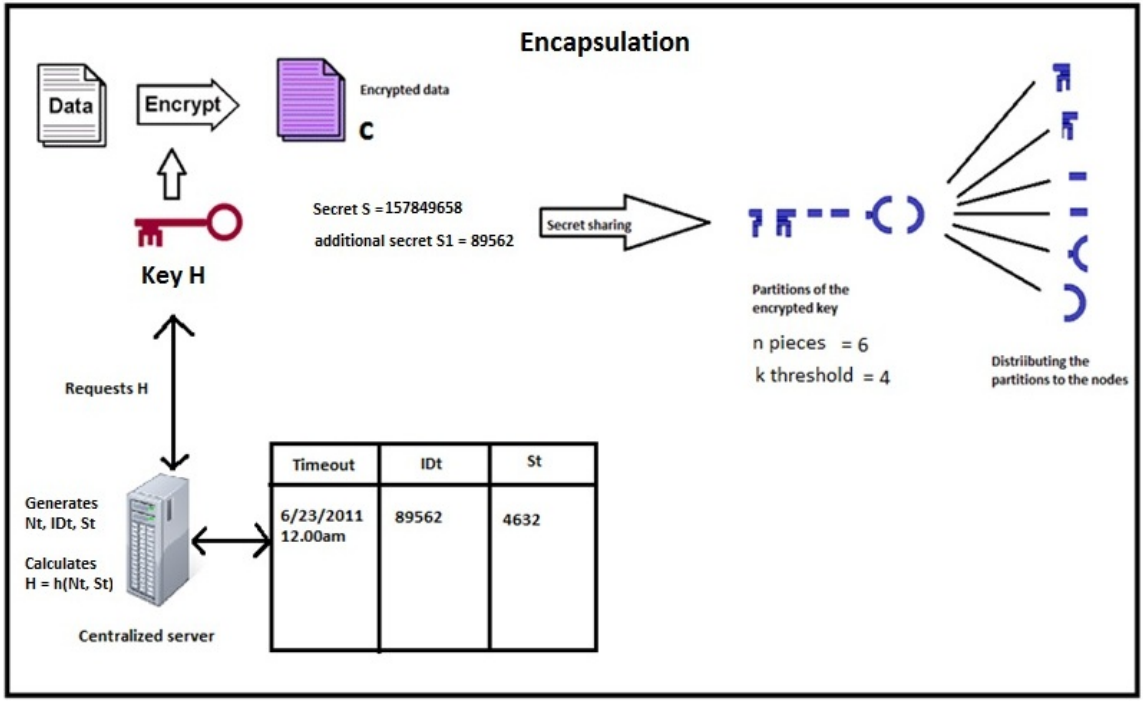

Figure 5: Proposed architecture encapsulation

3. The server sends $\{H_t, ID_t, N_t\}$ to Alice through a secured "SSL" channel and deletes $H$.
4. Alice encrypts *data* with key $H$ giving rise to $C$ the ciphertext and forgets $H$, now Alice has $(C, N_t, ID_t)$. It sends Ephemerizer an acknowledgment saying that the file is encrypted.
5. Ephemerizer then picks up an access key $L$ then use it in the cryptographic secure pseudorandom number generator to derive $n$ indices into DHT $I_1, I_2, I_3....I_n$.
6. Ephemerizer then performs the recursive secret sharing on $(N_t, ID_t)$, where secrets are $S = N_t$, $s_1 = ID_t$, $S$ is the main secret and $s_1$ are the additional secret messages. It generates $n$ shares $H_1, H_2, H_3....H_n$ where $k$ is the threshold.
7. Ephemerizer sprinkles $n$ shares $H_1, H_2, H_3....H_n$ at these pseudorandom locations $I_i$ in DHT. Then deletes $N_t$, and sends $L$ and $k$ to Alice.

8. The Ephemerizer stores the L values if the requested T is greater than DHT's fixed timeout, else it deletes it.
9. Finally Alice sends {L, C, k}
   L: Key to derive the indices related to that C.
   C: Cipher text.
   k: threshold minimum number of shares.

### 3.2 Decapsulation:

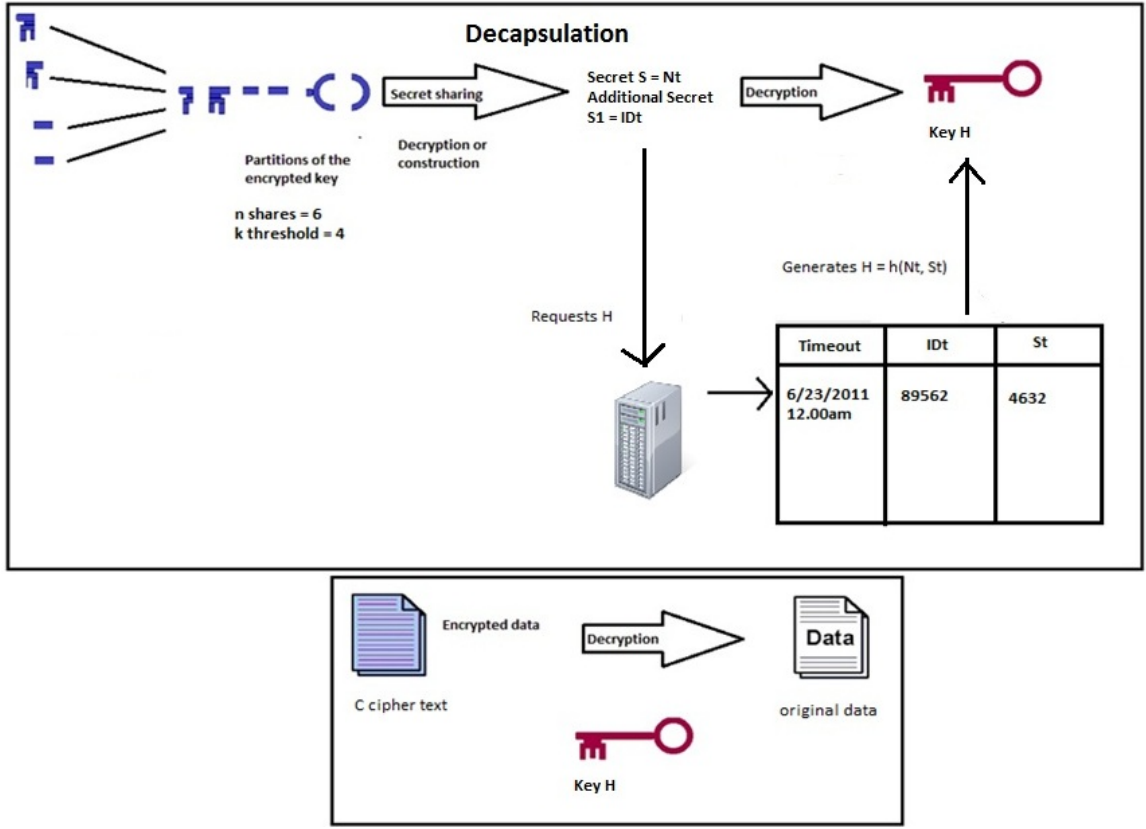

Figure 6: Proposed architecture decapsulations

1. On receiving {L, C, k}, Bob extracts the access key L. And derives the location of the shares. (Figure 6)
2. Bob retrieves the threshold k number of shares using the indices $I_1, I_2, I_3 .... I_n$, reconstructs the secret S and additional secrets S1. Where $S = N_t$, $s_1 = ID_t$.
3. When Bob wants to decrypt the cypher text C, he sends ($N_t$, $ID_t$) to the centralized server.
4. The centralized server checks if the $S_t$, $ID_t$ are valid or expired by checking the T. If it finds expired it sends an error message, else it finds St that is associated with the $ID_t$ that is still valid, calculates $H = h(N_t, S_t)$ and sends H to Bob through a secured SSL channel. And the server forgets the (T, $S_t$, $ID_t$) and H (if the code is designed to read only once).
5. On receiving H Bob decrypts the cyphertext C and gets the original *data*.

## 3.3 On Timeout:

For Vuze the timeout is 8 hour, and as per the churn effect key shares at nodes vanishes automatically after the timeout. If the timeout is less than 8 hours, Vanish will still retain key shares for eight hours. In the proposed architecture centralized server erases *IDt* and *St* from the table as shown in figure 7, making cipher text *C* completely unavailable to access, as key *H* to decrypt the cipher text *C* cannot be constructed without *H* and *St*.

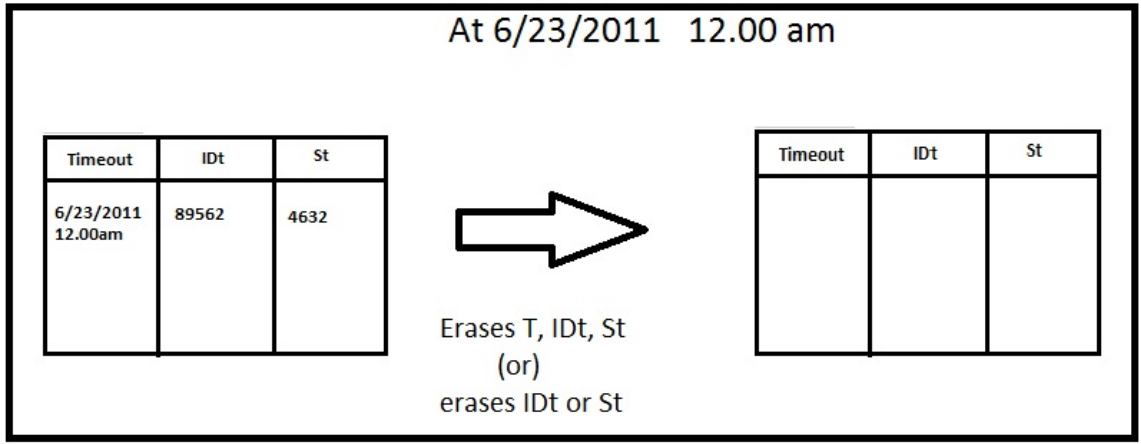

Figure 7: Proposed architecture timeout

## 3.4 Extending the lifetime of data:
Some users want their data to be available for more than 8 hours (considering the Vuze DHT). In such cases we use same technique as that of Vanish system, that is the refresh mechanism. Here just before timeout, Ephemerizer collects all the shares H1,H2,H3…Hk from the DHT and gets the secret S = Nt and s1 = IDt. It resplits them getting new shares and redistributes them with a new L.

## 3.5 Overcoming Drawbacks:
The presented architecture can handle most drawbacks in Vanish and in Ephemerizer architecture. This architecture can handle Sybil attacks too. Using this architecture a minimum timeout can be any number of hours specified by the user. If the specified timeout is 2 hours, after 2 hours centralized server will delete $ID_t$, $St$ from the table, which makes the key *H* unrecoverable. If the timeout specified is 8 hours, the keys are lost automatically by churn effect in DHT, unless life of encrypted data is extended. Even after 8 hours for Vuze, key shares are still available at the node, but they cannot be used to recover the key even though they are hacked. If the hacker was successfull in retrieving key shares, he still won't be able to reconstruct the original key.

As mentioned before, the Sybil attack works by continuously crawling DHT and saving each key share before timeout and makes a log caching all the key shares. But Sybil attacker does not know that the key *H* is encrypted, which cannot be acquired by just getting $N_t$ and $ID_t$. If Sybil attacker was successful in extracting all the additional secret messages and *Nt*, it still won't be able to access centralized server to get secret *St*.

## 4. SIMULATION AND RESULTS

### 4.1 Simulation
The simulation for the proposed method is done using Java 1.6 with extensive Java library functions. Here we have considered a 10 MB file as data *D*. Initially the user who holds data *D* performs data encryption. Here we use AES to encrypt the data using a 128 bit hash key *H*. The hash key *H* is derived from the Ephemerizer server, which is a simple multithreaded function that performs SHA-1 hashing on random numbers $(N_t, S_t)$ to generate hash key *H*. This hash key *H* along with additional secret $ID_t$ are secret shared using recursive secret sharing algorithm, which is programmed in Java. We distributed the shares into the nodes of the Open Chord DHT. The simulation is described in four sections.(figure 8)

1. Data encryption
2. Ephemerizer server
3. Recursive secret sharing
4. Distributed hash table

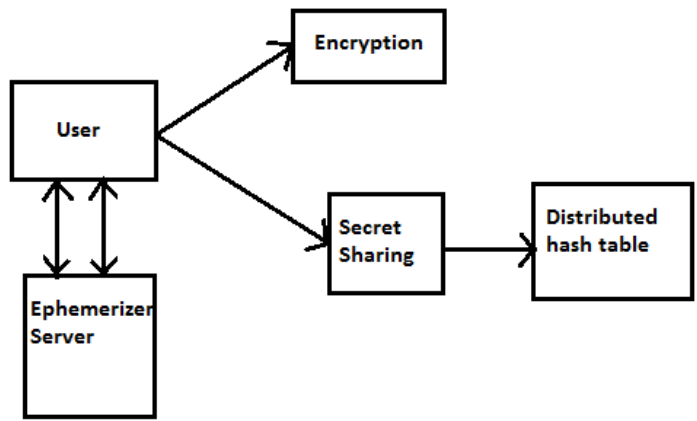

Figure 8: Proposed architecture block diagram

### 4.1.1 Data Encryption
The proposed architecture is designed in a way that, it can be used in Cloud, external server storage and to transfer data securely. Data is encrypted using the AES encryption technique.
**Advanced Encryption Standard** (**AES**):
　　AES is a federal standard for private-key or symmetric cryptography, used for the encryption of electronic data. It was first introduced by the U.S. government, now used word wide. There are many packages of AES encryption. AES was used by the National Security Agency (NSA) for storing top secret data. AES was originally called Rijndael, the cipher was developed by two Belgian cryptographers, J. Daemen and V. Rijmen.

AES is based on substitution permutation network as the design principle. It works efficiently and fast in both software and hardware. AES has a fixed block size of 128 bits and a key size of 128, 192, or 256 bits. The block size of the AES has a maximum limit of 256 bits, but the key size has no maximum limits. AES operates on a 4×4 column-major order matrix of bytes, termed the *state*. Most of the AES calculations are performed in a special finite field.

**AES in Java**

The AES standard has been incorporated into several Java technology offerings. Beginning with Java 2 SDK (software development kit), Standard Edition (J2SE) v 1.4.0, JCE (Java cryptography extension) [18] was integrated with the SDK and JRE (Java runtime environment). JCE provides framework and implementation for encryption, key generation and agreement. JCE supplements the Java platform, which already includes interfaces and implementations of message digests and digital signatures.

### 4.1.2 Ephemerizer server

As discussed earlier, Ephemerizer is simple server external to the user. It performs simple operations like creation, deletion and distribution of keys. In this simulation, we have considered Ephemerizer server as a multithreaded function that performs all the above operations. The Java threading makes it look like a server that runs constantly in a loop. Each thread performs a separate task. One thread performs the creation and distribution of keys while the other checks the $T$- timeout periodically. If the thread finds $T$ expired, it simply deletes $S_t$. All the variables *ID, $S_t$, T* are stored in JTables (Java Tables).

**Hashing SHA-1:**

In this simulation we have used SHA-1 hashing technique in Java. This hashing technique is used in calculating the hash key $H$, $H = (N_t, S_t)$. SHA-1 is a cryptographic message digest algorithm similar to MD5. SHA-1 hash is considered to be one of the most secure hashing functions, producing a 160-bit digest from any data with a maximum size of $2^{64}$ bits. We used Java built in classes to compute SHA-1 hash.

### 4.1.3 Recursive Secret Sharing

The recursive secret sharing was programed using simple Java programming,
In this simulation Lagrange's interpolation was performed using Aitkens method [19]. We have recursively secret shared the key $N_t$ (nonce) of size 128 bit and $ID_t$ 16 bit to generate various numbers of shares. For efficient results, we have considered the threshold ratio of 100%.

### 4.1.4 DHT:

In this simulation we used Open Chord as the DHT. Open Chord is an open source implementation of Chord distributed hash table using Java-based implementation of Chord DHT. Open Chord provides the possibility to use Chord distributed hash table within Java applications by providing an API to store all serializable Java objects within the distributed hash table.

It provides an interface for Java applications to take part as a peer within a DHT and to store and retrieve arbitrary data from this DHT. Open Chord is called open, as it is distributed under GNU General Public License (GPL), so that it can be used and extended for own purposes for free as desired. The Open Chord libraries are extracted from Open Chord package zip file [20].

**Important Features of the Open Chord**
- It can store any serializable JO (Java object) within the DHT.
- It can create own key implementations used along with DHT by implementing an interface of Open Chord API.
- Facilitates configurable replication of entries in DHT.
- Provides two protocols for communicating in between the chord nodes:
    - Local method calls: This protocol is used to create a DHT within one JVM (Java Virtual Machine) for testing and visualization purposes.
    - Java Sockets: This protocol creates a DHT, distributed over different nodes (JVMs).

We used *local method call* protocol for communicating between 200 chord nodes.

The basic operations of a DHT are *Store(key, val), val = Retrieve(key)*. A *key* value controls which node(s) stores the value *val*. Each node is responsible for some section of the space. In Open Chord these operations are performed using the following methods
*public void retrieve ( Key key , ChordCallback callback );*
*public void insert ( Key key , Serializable entry , ChordCallback callback );*
A new network can be created with help of the methods *create(), create(URL localURL)*, and *create(URL localURL, ID localID)*. The join methods allow a peer to join an existing Open Chord network
*public void join ( URL localURL , URL bootstrapURL )*
 *throws ServiceException ;*

## 4.2 Results:
We measure the performance of the new architecture by measuring time elapsed to encapsulate and decapsulate. Our main purpose is to measure the time and determine whether the new system is fast enough for daily usage. In simulation, we have used a Intel Core 2 Duo with 2.00GHz processor speed and 4 GB RAM. We measure the time of encapsulation and decapsulation, considering a 10 MB files with various shares. In this simulation we have considered a threshold ratio of 100%.
We executed encapsulation and de-capsulation operations and measured the time spent in the four main runtime components: DHT operation (storing and retrieving shares), recursive secret sharing operation (splitting and combining shares), Ephemerizer server (storing and creating hash keys), and encryption/decryption operation. We observed that the DHT component accounted for over 99% of execution time for encapsulation and de-capsulation operations on data of small and medium size. For data of much larger size, encryption and decryption became a dominant component. For recursive secret sharing, we tried various other Lagrange's interpolation methods like the upwards/downward correction method [21] and Apaches Lagrange's interpolation classes [22], and they yielded same results.

Executing DHT component using local call protocol, which created the DHT within the system's Java Virtual Machine (JVM), we observed that systems with different configurations exhibit different results. Figure 10, 11 shows operation time scale with different number of shares for a fixed threshold ration of 100%. Scaling with n (number of shares) is important, as data's security relies on this parameter. Figure 11 shows that time for encapsulation and de-capsulation grows linearly with the number of shares. De-capsulation took somewhat less time than encapsulation, because of DHT. Based on these results, we believe that parameters of n = 50 and the threshold of 90% provide an excellent tradeoff for security and performance.

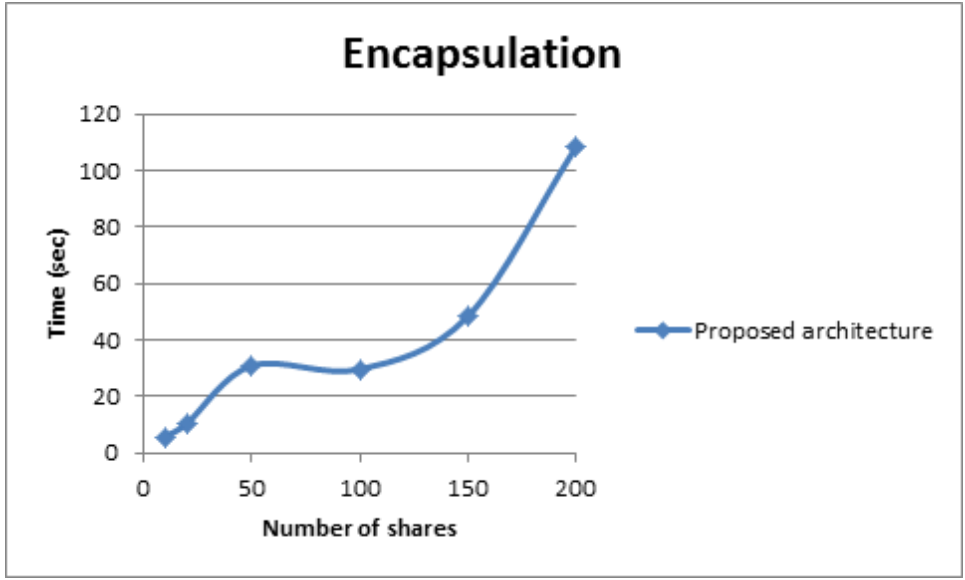

Figure 9: Encapsulation time performance graph

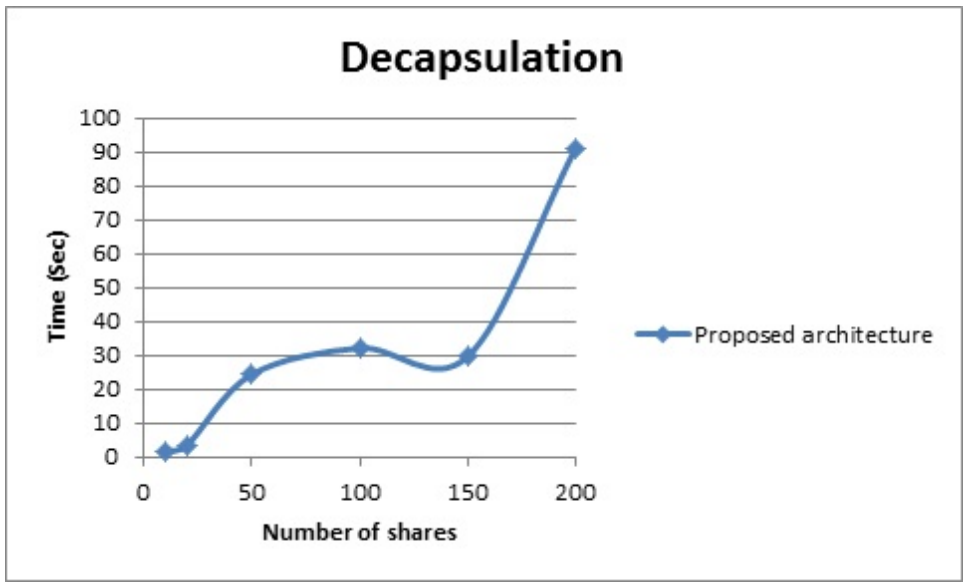

Figure 10: Decapsulation time performance graph

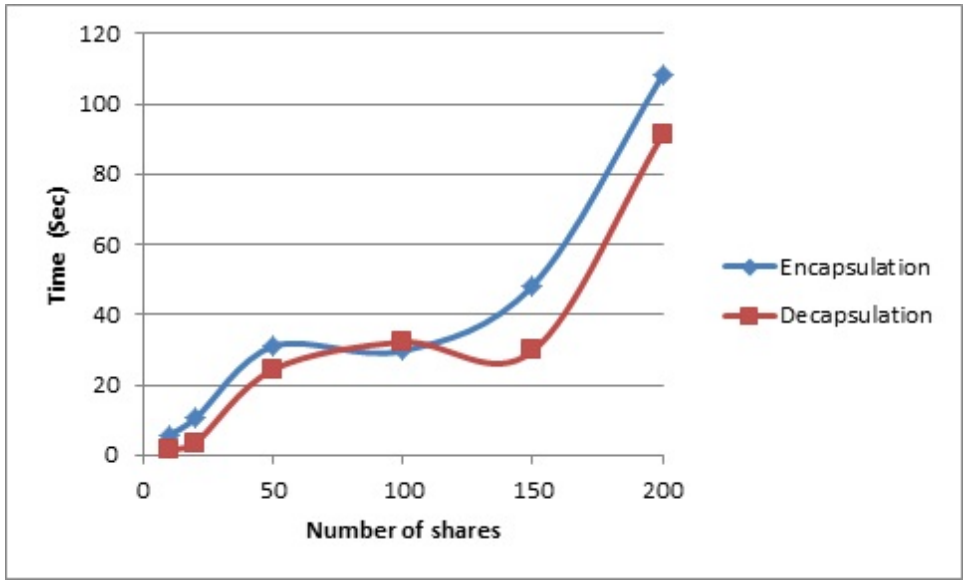

Figure.11: Encapsulation and decapsulation performance graph

| N | Time (sec) | |
|---|---|---|
| | Encapsulation | Decapsulation |
| 10 | 5.78 | 1.68 |
| 20 | 10.56 | 3.56 |
| 50 | 31.043 | 24.444 |
| 100 | 29.719 | 32.183 |
| 150 | 48.394 | 29.74 |
| 200 | 108.187 | 91.178 |

Figure.12: Encapsulation and decapsulation execution time

**CONCLUSION**

In this paper I have designed a new architecture to enhance data security by making the data disappear after a time limit. This was done by simply encrypting the data with a key and later making the key disappear after timeout, ultimately making the data inaccessible. We consider key management in centralized as well as decentralized environments. We have successfully implemented this architecture.

This system is more secure and efficient than the earlier system called Vanish. This system can be used for storing data on the Cloud or on other servers, or in a database.